\documentclass[11pt]{article}
\usepackage[latin9]{inputenc}
\usepackage{amsmath}
\usepackage{amssymb}
\usepackage{esint}
\usepackage[unicode=true,
 bookmarks=false,
 breaklinks=false,pdfborder={0 0 1},backref=section,colorlinks=false]
 {hyperref}

\makeatletter

\providecommand{\tabularnewline}{\\}

\@ifundefined{date}{}{\date{}}

\usepackage{esint}
\usepackage{latexsym}
\usepackage{graphicx}\usepackage{bm}\usepackage{longtable}

\usepackage{xcolor}

\@addtoreset{equation}{section}

\makeatother

\begin{document}
\title{The theta-dependent Yang-Mills theory at finite temperature in a holographic
description}
\maketitle
\begin{center}
\footnote{Email: siwenli@dlmu.edu.cn}Si-wen Li\emph{$^{\dagger}$}
\par\end{center}

\begin{center}
\emph{$^{\dagger}$Department of Physics, School of Science,}\\
\emph{Dalian Maritime University, }\\
\emph{Dalian 116026, China}\\
\par\end{center}

\vspace{8mm}

\begin{abstract}
The theta-dependent gauge theories can be studied by using holographic
duality through string theory on certain spacetimes. Via this correspondence
we consider a stack of $N_{0}$ dynamical D0-branes as D-instantons
in the background sourced by $N_{c}$ coincident non-extreme black
D4-branes. According to the gauge-gravity duality this D0-D4 brane
system corresponds to Yang-Mills theory with a theta angle at finite
temperature. We solve the IIA supergravity action by taking account
into a sufficiently small backreaction of the D-instantons and obtain
an analytical solution for our D0-D4-brane configuration. Then the
dual theory in the large $N_{c}$ limit can be holographically investigated
with the gravity solution. In the dual field theory, we find the coupling
constant exhibits the property of asymptotic freedom as it is expected
in QCD. The contribution of the theta-dependence to the free energy
gets suppressed at high temperature which is basically consistent
with the calculation by using the Yang-Mills instanton. The topological
susceptibility in the large $N_{c}$ limit vanishes and this behavior
remarkably agrees with the implications from the simulation results
at finite temperature. Besides we finally find a geometrical interpretation
of the theta-dependence in this holographic system.
\end{abstract}
\newpage{}

\tableofcontents{}

\section{Introduction}

The spontaneous CP violation in Quantum Chromodynamics (QCD) has been
studied for very long time and such effects can usually be described
by introducing a $\theta$ term to the four dimensional (4d) action
for the gauge theories as \cite{key-1},

\begin{equation}
S=-\frac{1}{2g_{YM}^{2}}\mathrm{Tr}\int F\wedge^{*}F+i\frac{\theta}{8\pi^{2}}\mathrm{Tr}\int F\wedge F,\label{eq:1}
\end{equation}
where $g_{YM}$ is Yang-Mills coupling constant and the second term
defines the topological charge density with a $\theta$ angle. While
the experimental value of the theta angle is stringently small $\left(\left|\theta\right|\leq10^{-10}\right)$,
the dependence of Yang-Mills theory and QCD on theta attracts great
theoretical and phenomenological interests, e.g. the study of the
large $N_{c}$ behavior \cite{key-2}, the glueball spectrum \cite{key-3},
the deconfinement phase transition \cite{key-4,key-5} and the Schwinger
effect \cite{key-6}. Particularly there is an open question in hadron
physics, that is whether a theta vacua can be created in hot QCD.
To figure out this issue, there have been some progresses achieved
in \cite{key-7,key-8,key-9,key-10,key-11,key-12} and one of the most
famous proposal is to search for the chiral magnet effect (CME) in
heavy-ion collisions \cite{key-13,key-14,key-15,key-16} to confirm
the theta dependence at high temperature.

On other hand the AdS/CFT correspondence, or more generally the gauge-gravity
(string) duality, has rapidly become a powerful tool to investigate
the strongly coupled quantum field theory (QFT) since 1997 \cite{key-17,key-18,key-19}.
In the holographic approach to study QCD or Yang-Mills theory, a concrete
model is proposed by Witten \cite{key-20} and Sakai and Sugimoto
\cite{key-21,key-22} (named as WSS model) based on the IIA string
theory. This model is significantly successful since it almost includes
all necessary ingredients of QCD or Yang-Mills theory in a very simple
way, e.g. the fundamental quarks and mesons \cite{key-21,key-22,key-23},
baryon \cite{key-24,key-25}, the phase diagram of hot QCD \cite{key-26,key-27,key-28,key-29,key-30},
glueball spectrum \cite{key-31,key-32} and the interactions of hadrons
\cite{key-33,key-34,key-35,key-36,key-37,key-38}. Due to the non-perturbative
properties of the theta dependence, it has been recognized that the
D-branes as D-instantons in the bulk geometry plays the role of the
theta angle in the dual theory \cite{key-39,key-40,key-41}. Via this
viewpoint, the holographic correspondence of theta-dependence in QCD
or Yang-Mills theory has been systematically studied by using the
WSS model with D0-branes as D-instantons at zero temperature, or without
temperature in \cite{key-42,key-43,key-44,key-45,key-46,key-47,key-48,key-49,key-50}.

\begin{table}[h]
\begin{centering}
\begin{tabular}{|c|c|c|c|c|c|c|c|c|c|c|}
\hline 
 & 0 & 1 & 2 & 3 & 4 & 5$\left(\rho\right)$ & 6 & 7 & 8 & 9\tabularnewline
\hline 
\hline 
$N_{0}$ smeared D0-branes & = & = & = & = & - &  &  &  &  & \tabularnewline
\hline 
$N_{c}$ black D4-branes & - & - & - & - & - &  &  &  &  & \tabularnewline
\hline 
\end{tabular}
\par\end{centering}
\caption{\label{tab:1}The configuration of $N_{0}$ smeared D0 and $N_{c}$
black D4-branes with the compactified direction $x^{4}$. The ``-''
represents that the D-branes extend along this direction and ``=''
represents the direction where the D0-branes are smeared.}
\end{table}

To analyze the theta dependence at finite temperature, there have
been various researches by using simulations and the results imply
some large $N_{c}$ behaviors are different from the situations of
zero temperature or without temperature \cite{key-1}. In the current
status of the holographic approaches, the theta dependence at finite
temperature is studied mostly in the $\mathcal{N}=4$ super Yang-Mills
theory by the D(-1)-D3 brane configuration e.g. \cite{key-39,key-51,key-52}.
In the opposite, few lectures discuss specifically QCD or Yang-Mills
theory at finite temperature through the D0-D4 brane configuration.
So we are motivated to fill this blank by exploring a way to combine
the theta-dependent Yang-Mills at finite temperature with the IIA
string theory. In our setup, we adopt the gravity background sourced
by a stack of $N_{c}$ black non-extreme D4-branes since the dual
field theory in this background exhibits deconfinement at finite temperature
\cite{key-26}. Then we introduce $N_{0}$ coincident D0-branes as
D-instantons into the D4-brane background by taking account into a
very small backreaction to the bulk geometry. Hence the D-instantons
are dynamical and this setup is coincident with the bubble D0-D4 configuration
in \cite{key-42,key-43,key-44,key-45,key-46,key-47,key-48,key-49,key-50}\footnote{Here we emphasize that the dual theory in the approach of the bubble
D0-D4 configuration is defined at zero-temperature limit, or defined
without a concrete temperature according to \cite{key-26,key-29,key-30}.
The dual theory includes a finite temperature is the distinguishing
feature in our setup.}. In order to search for an analytical supergravity solution, we further
assume that the D0-branes are homogeneously smeared in the worldvolume
of the D4-branes and this D-brane configuration is illustrated in
Table \ref{tab:1}. Solving the effective 1d gravity action, we indeed
obtain a particularly analytical solution, then we examine the coupling
constants and renormalized ground-state energy by the gravity solution.
The coupling constant indicates the property of asymptotic freedom
and the free energy gets suppressed at high temperature. Besides the
topological susceptibility in the large $N_{c}$ limit vanishes. We
find that all these results agree with the implications of simulation
reviewed in \cite{key-1}, or the well-known properties of QCD and
Yang-Mills theory, thus our work might offer a holographic way to
study the issues proposed in \cite{key-7,key-8,key-9,key-10,key-11,key-12,key-13,key-14,key-15}.

Here is the outline of this manuscript. In Section 2, we first discuss
the general formulas of the black D0-D4 system based on the IIA supergravity.
Then comparing them with the black D4-brane solution, we obtain a
particular solution by including some physical constraints. In Section
3, we evaluate the coupling constant and free energy density by our
gravity solution. We also give an geometric interpretion of the theta-dependence
in this D0-D4 system. The final Section is the summary and discussion.
Our gravity solution expressed in the $U$ coordinate is summarized
in the Appendix.

\section{The supergravity description}

\subsection{General setup}

In this section, let us explore the holographic description based
on the $N_{0}$ D0- and $N_{c}$ D4-branes with the configuration
illustrated in Table \ref{tab:1}. As the gauge-gravity duality is
valid in the large $N_{c}$ limit, we can first define the 4d 't Hooft
coupling as $\lambda_{4}=g_{YM}^{2}N_{c}$ where $g_{YM}$ is the
Yang-Mills coupling and $\lambda_{4}$ is fixed when $N_{c}\rightarrow\infty$.
Then in order to take into account a small backreaction of the $N_{0}$
D0-branes, we further require $N_{0}\rightarrow\infty$ while

\begin{equation}
\frac{N_{0}}{N_{c}}=\mathcal{C}\ \mathrm{fixed},\ \mathcal{C}\ll1.\label{eq:2}
\end{equation}

\noindent Here $\mathcal{C}$ is a fixed constant in the limitation
of $N_{c},N_{0}\rightarrow\infty$ and we note that this limit is
similar as the Veneziano limit discussed in \cite{key-29,key-30}.
Keeping this in mind, let us consider the dynamics of the 10d bulk
geometry which is described by the type IIA supergravity. In string
frame the action is given as,

\begin{equation}
S_{IIA}=\frac{1}{2\kappa_{0}^{2}}\int d^{10}x\sqrt{-g}\left[e^{-2\phi}\left(\mathcal{R}+4\left(\partial\phi\right)^{2}\right)-\frac{1}{2}\left|F_{2}\right|^{2}-\frac{1}{2}\left|F_{4}\right|^{2}\right],\label{eq:3}
\end{equation}
where $2\kappa_{0}^{2}=\left(2\pi\right)^{7}l_{s}^{8}$ and $l_{s}=\sqrt{\alpha^{\prime}}$
is the string length. $F_{4}=dC_{3},F_{2}=dC_{1}$ is the Ramond-Ramond
four and two form sourced by the $N_{c}$ D4-branes and $N_{0}$ D0-branes.
We have used $\mathcal{R}$ and $\phi$ to denote the 10d scalar curvature
and the dilaton field respectively. Since the D0-branes as D-instantons
are extended along $x^{4}$ direction and homogeneously smeared in
the directions of $\left\{ x^{0},x^{i}\right\} ,i=1,2,3$, we may
search for a possible solution by using the metric ansatz written
as \cite{key-26,key-29,key-30},

\begin{align}
ds^{2} & =-e^{2\widetilde{\lambda}}dt^{2}+e^{2\lambda}\delta_{ij}dx^{i}dx^{j}+e^{2\lambda_{s}}\left(dx^{4}\right)^{2}+l_{s}^{2}e^{-2\varphi}d\rho^{2}+l_{s}^{2}e^{2\nu}d\Omega_{4}^{2}.\label{eq:4}
\end{align}
The the Ramond-Ramond $C_{1}$ form and its field strength $F_{2}$
is assumed to be,

\begin{align}
C_{1} & =\left[h\left(\rho\right)+H\right]dx^{4},\nonumber \\
F_{2} & =dC_{1}=\partial_{\rho}hdx^{4}\wedge d\rho,\label{eq:5}
\end{align}
where $H$ is a constant and $h\left(\rho\right)$ is a function to
be solved. To find a static and homogeneous solution by the ansatz
(\ref{eq:4}), we further assume that the functions $\widetilde{\lambda},\lambda,\lambda_{s},\varphi,\nu$
and the dilaton $\phi$ only depend on the holographic coordinate
$\rho$. Hence the action (\ref{eq:3}) could be rewritten as an effective
1d action by inserting (\ref{eq:4}) (\ref{eq:5}) into (\ref{eq:3})
which leads to,

\begin{equation}
S_{IIA}=\mathcal{V}\int d\rho\left[-3\dot{\lambda}^{2}-\dot{\lambda}_{s}^{2}-\dot{\widetilde{\lambda}}^{2}-4\dot{\nu}^{2}+\dot{\varphi}^{2}-\frac{1}{2}e^{3\lambda+\widetilde{\lambda}-\lambda_{s}+4\nu+\varphi}\dot{h}^{2}+V+\mathrm{total\ derivative}\right].\label{eq:6}
\end{equation}
We have used ``.'' to represent derivatives which are w.r.t. $\rho$
and,

\begin{align}
V & =12e^{-2\nu-2\varphi}-Q_{c}^{2}e^{3\lambda+\lambda_{s}+\widetilde{\lambda}-4\nu-\varphi},\ \mathcal{V}=\frac{1}{2k_{0}^{2}}V_{3}V_{S^{4}}\beta_{T}\beta_{4}l_{s}^{3},\nonumber \\
\varphi & =2\phi-3\lambda-\widetilde{\lambda}-\lambda_{s}-4\nu,\ \ Q_{c}=\frac{3\pi^{2}l_{s}}{\sqrt{2}\kappa_{0}}\int_{S^{4}}F_{4}.
\end{align}
Here $\beta_{4},\beta_{T}$ refers to the size of (time) $x^{0}$
and $x^{4}$ direction\footnote{Since we would consider a 4d dual field theory at finite temperature,
the $x^{0}$ and $x^{4}$ directions have to be compactified on $S^{1}$
as discussed in \cite{key-26,key-27,key-28,key-29,key-30} in this
model. And $\beta_{4},\beta_{T}\rightarrow\infty$ corresponds to
the decompactified limit.}, $V_{3}$ represents the 3d spacial volume and $V_{S^{4}}=\frac{8\pi^{2}}{3}$
is the volume of a unit $S^{4}$. Then the solution for $C_{1}$ may
immediately be obtained as,

\begin{equation}
\dot{h}\left(\rho\right)=-q_{\theta}e^{2\lambda_{s}-2\phi},\label{eq:7}
\end{equation}
where $q_{\theta}$ is an integration constant related to the $\theta$
angle and this would become more clear later. Note that the 1d action
(\ref{eq:6}) has to be supported by the following zero-energy constraint
\cite{key-29,key-30,key-45},

\begin{equation}
-3\dot{\lambda}^{2}-\dot{\lambda}_{s}^{2}-\dot{\widetilde{\lambda}}^{2}-4\dot{\nu}^{2}+\dot{\varphi}^{2}-\frac{1}{2}e^{3\lambda+\widetilde{\lambda}-\lambda_{s}+4\nu+\varphi}\dot{h}^{2}-V=0,\label{eq:8}
\end{equation}
so that the equations of motion from the 1d effective action (\ref{eq:6})
would be coincident with those from the 10d action (\ref{eq:3}) if
the homogeneous ansatz (\ref{eq:4}) is adopted.

Afterwards the full equations of motion could be obtained by varying
the 1d action (\ref{eq:6}) which are given as,

\begin{eqnarray}
\ddot{\lambda}-\frac{Q_{c}^{2}}{2}e^{6\lambda+2\lambda_{s}+2\widetilde{\lambda}-2\phi} & = & \frac{q_{\theta}^{2}}{4}e^{2\lambda_{s}-2\phi},\nonumber \\
\ddot{\lambda}_{s}-\frac{Q_{c}^{2}}{2}e^{6\lambda+2\lambda_{s}+2\widetilde{\lambda}-2\phi} & = & -\frac{q_{\theta}^{2}}{4}e^{2\lambda_{s}-2\phi},\nonumber \\
\ddot{\widetilde{\lambda}}-\frac{Q_{c}^{2}}{2}e^{6\lambda+2\lambda_{s}+2\widetilde{\lambda}-2\phi} & = & \frac{q_{\theta}^{2}}{4}e^{2\lambda_{s}-2\phi}\nonumber \\
\ddot{\nu}+\frac{Q_{c}^{2}}{2}e^{6\lambda+2\lambda_{s}+2\widetilde{\lambda}-2\phi}-3e^{6\lambda+2\lambda_{s}+2\widetilde{\lambda}-4\phi+6\nu} & = & \frac{q_{\theta}^{2}}{4}e^{2\lambda_{s}-2\phi},\nonumber \\
\ddot{\phi}-\frac{Q_{c}^{2}}{2}e^{6\lambda+2\lambda_{s}+2\widetilde{\lambda}-2\phi} & = & \frac{3q_{\theta}^{2}}{4}e^{2\lambda_{s}-2\phi}.\label{eq:10}
\end{eqnarray}
In order to find a solution for (\ref{eq:10}), let us introduce the
new variables $\gamma,p,\chi$ defined as,

\begin{equation}
\gamma=6\lambda+2\lambda_{s}+2\widetilde{\lambda}-2\phi,\ p=6\lambda+2\lambda_{s}+2\widetilde{\lambda}-4\phi+6\nu,\ \chi=2\lambda_{s}-2\phi.\label{eq:11}
\end{equation}
So the (\ref{eq:10}) reduces to three simple equations,

\begin{equation}
\ddot{\gamma}-4Q_{c}^{2}e^{\gamma}=0,\ \ddot{p}-18e^{p}=0,\ \ddot{\chi}+2q_{\theta}^{2}e^{\chi}=0.\label{eq:12}
\end{equation}
 And the solution for the equations in (\ref{eq:12}) could be analytically
obtained as,

\begin{align}
\gamma & =-2\log\left[a_{1}-e^{-a_{2}\rho}\right]-a_{2}\rho+\log\left[\frac{a_{1}a_{2}^{2}}{2Q_{c}^{2}}\right],\nonumber \\
p & =-2\log\left[a_{3}-e^{-a_{4}\rho}\right]-a_{4}\rho+\log\left[\frac{a_{3}a_{4}^{2}}{9}\right],\nonumber \\
\chi & =-2\log\left[a_{5}+e^{-a_{6}\rho}\right]-a_{6}\rho+\log\left[\frac{a_{5}a_{6}^{2}}{q_{\theta}^{2}}\right],\label{eq:13}
\end{align}
where $a_{1,2,3,4,5,6}$ are integration constants. According to (\ref{eq:10}),
we on the other hand have,

\begin{align}
\lambda-\widetilde{\lambda} & =b_{1}\rho+b_{2}\nonumber \\
\lambda-\lambda_{s} & -\phi+\widetilde{\lambda}=b_{3}\rho+b_{4},\label{eq:14}
\end{align}
where $b_{1,2,3,4}$ are additional integration constants. Altogether
with (\ref{eq:13}) and (\ref{eq:14}), we could obtain the full solution
for (\ref{eq:4}) as,

\begin{align}
\lambda & =\frac{1}{8}\left(\gamma-\chi\right)+\frac{1}{4}\left(b_{2}+b_{1}\rho\right),\nonumber \\
\lambda_{s} & =\frac{1}{8}\left(\gamma+\chi\right)-\left(\frac{b_{1}}{4}+\frac{b_{3}}{2}\right)\rho-\frac{b_{2}}{4}-\frac{b_{4}}{2},\nonumber \\
\widetilde{\lambda} & =\frac{1}{8}\left(\gamma-\chi\right)-\frac{3b_{2}}{4}-\frac{3b_{1}}{4}\rho,\nonumber \\
\phi & =\frac{1}{8}\left(\gamma-3\chi\right)-\left(\frac{b_{1}}{4}+\frac{b_{3}}{2}\right)\rho-\frac{b_{2}}{4}-\frac{b_{4}}{2},\nonumber \\
\nu & =\frac{p}{6}-\frac{1}{8}\left(\gamma+\chi\right)-\left(\frac{b_{1}}{12}+\frac{b_{3}}{6}\right)\rho-\frac{b_{2}}{12}-\frac{b_{4}}{6}.\label{eq:15}
\end{align}
Besides, the zero-energy constraint (\ref{eq:8}) reduces to the following
relation,

\begin{equation}
-3a_{2}^{2}+8a_{4}^{2}-3a_{6}^{2}-20b_{1}^{2}-8b_{1}b_{3}-8b_{3}^{2}=0.
\end{equation}
While all the integration constants should be further determined by
some additional physical conditions, we note that these integration
constants could depend on $q_{\theta}$ which is the only parameter
in the solution.

\subsection{A particular solution}

In this section, let us discuss a particular solution to fix the integration
constants in the supergravity solution obtained in the last section.
Since $\left|\theta\right|$ is usually very small in Yang-Mills theory,
we consider a sufficiently small backreaction of the D-instantons
(D0-branes) in the black D4 configuration. Therefore we require the
solution (\ref{eq:15}) must be able to return to the pure black D4-brane
solution if $q_{\theta}\rightarrow0$ i.e. no D0-branes. Hence the
black D4-brane solution corresponds to the situation of $C_{1}=0$\footnote{Strictly speaking, the black D4-brane solution corresponds to the
situation that $C_{1}$ is a constant because the IIA action (\ref{eq:3})
is invariant under the gauge transformation $C_{1}\rightarrow C_{1}+d\Lambda$
where $\Lambda$ is an arbitrary function. Thus we can choose a particular
gauge condition so that $C_{1}=0$ corresponds to the situation of
the black D4-brane solution.} in (\ref{eq:3}) and in the near-horizon limit the solution is given
as,

\begin{align}
 & ds^{2}=\left(\frac{U}{R}\right)^{3/2}\left[-f_{T}\left(U\right)dt^{2}+\delta_{ij}dx^{i}dx^{j}+\left(dx^{4}\right)^{2}\right]+\left(\frac{R}{U}\right)^{3/2}\left[\frac{dU^{2}}{f_{T}\left(U\right)}+U^{2}d\Omega_{4}^{2}\right],\nonumber \\
 & f_{T}\left(U\right)=1-\frac{U_{T}^{3}}{U^{3}},\ e^{\phi}=g_{s}\left(\frac{U}{R}\right)^{3/4},\ F_{4}=3R^{3}g_{s}^{-1}\omega_{4},\ R^{3}=\pi g_{s}N_{c}l_{s}^{3},\label{eq:17}
\end{align}
where $g_{s},\omega_{4}$ represents the string coupling constant
and the volume form of $S^{4}$. Accordingly we identify the solution
(\ref{eq:17}) as the zero-th order solution of (\ref{eq:13}) and
rewrite it in terms of $\gamma,p,\chi$ defined as in (\ref{eq:11}),

\begin{align}
\gamma_{0} & =-2\log\left[1-e^{-3a\rho}\right]-3a\rho+\log\left[\frac{U_{T}^{6}}{g_{s}^{2}R^{6}}\right],\nonumber \\
p_{0} & =-2\log\left[1-e^{-3a\rho}\right]-3a\rho+\log\left[\frac{U_{T}^{6}}{g_{s}^{4}l_{s}^{6}}\right],\nonumber \\
\chi_{0} & =-2\log\left[g_{s}\right].\label{eq:18}
\end{align}
This gives the relation of $\rho$ and the usually used $U$ coordinate
in (\ref{eq:17}) as,

\begin{equation}
\rho=-\frac{b_{\theta}}{3a}\log\left[1-\frac{U_{T}^{3}}{U^{3}}\right],\ a=\frac{\sqrt{2}Q_{c}U_{T}^{3}}{3R^{3}g_{s}}=\frac{U_{T}^{3}}{l_{s}^{3}g_{s}^{2}},\ Q_{c}=\frac{3\pi N_{c}}{\sqrt{2}}.
\end{equation}
Here $b_{\theta}$ is another constant depended on $\theta$ which
is required as $b_{\theta}\rightarrow1$ if $q_{\theta}\rightarrow0$.
Comparing (\ref{eq:18}) with (\ref{eq:13}), it implies that in the
limitation of $q_{\theta}\rightarrow0$ there must be $a_{1,3}\rightarrow1,a_{2,4}\rightarrow3a,a_{5}a_{6}^{2}\rightarrow\frac{4q_{\theta}^{2}}{g_{s}^{2}},a_{6}\rightarrow q_{\theta}$
so that $\gamma,p,\chi$ returns to $\gamma_{0},p_{0},\chi_{0}$ consistently.
In this sense, we could in particular choose $a_{5}=1,a_{6}=2\left|q_{\theta}\right|g_{s}^{-1}$
so that $a_{1}=a_{3}=1,a_{2}=3a,b_{2}=0,b_{4}=-\log\left[g_{s}\right]$
as the most simply solution. Moreover we require that $g_{00}\sim\tilde{\lambda},g_{ij}\sim\lambda,g_{\Omega\Omega}\sim\nu$
has to behave as same as they are in the zero-th order solution (\ref{eq:17})
in the IR region (i.e. $U\rightarrow U_{T}$, $\rho\rightarrow\infty$)
so that the holographic duality constructed on the $N_{c}$ D4-branes
basically remains in the low-energy theory. Therefore we have the
following relations,

\begin{equation}
b_{1}=\frac{1}{2}\left(a_{2}-a_{6}\right),\ a_{4}=\frac{a_{2}^{2}+a_{6}^{2}}{a_{2}+a_{6}}.
\end{equation}
On the other hand the zero-energy constraint (\ref{eq:8}) reduces
to an extra relation to determine $b_{3}$ which is,

\begin{equation}
b_{3}=-\frac{1}{2}b_{1}-\frac{\sqrt{2}}{4}\sqrt{-3a_{2}^{2}+8a_{4}^{2}-3a_{6}^{2}-18b_{1}^{2}},\ \ -3a_{2}^{2}+8a_{4}^{2}-3a_{6}^{2}-18b_{1}^{2}\geq0.
\end{equation}
The above constraints imply that our solution would be valid only
if $\left|q_{\theta}\right|\leq\frac{3}{2}ag_{s}$ and it is consistent
with our assumption that the backreaction of D-instantons is sufficiently
small. The constant $b_{\theta}$ could be determined by additionally
requiring that $g_{UU}\sim\psi$ behaves as same as it in (\ref{eq:17})
at $U=U_{T}$ and this gives 
\begin{equation}
b_{\theta}=\frac{9a^{2}g_{s}^{2}-6ag_{s}q_{\theta}}{9a^{2}g_{s}^{2}+4q_{\theta}^{2}}.
\end{equation}
For the reader's convenience, we have summarized the current solution
in the $U$ coordinate in the Appendix and one could compare it with
the zero-th order solution (\ref{eq:17}) directly. Notice that our
solution also has the same behaves as (\ref{eq:17}) in the UV region
(i.e. $U\rightarrow\infty$, $\rho\rightarrow0$).

\section{The dual field theory}

\subsection{The running coupling}

To start this section, let us examine the dual field theory interpretation
of the above supergravity solution in Section 2.2 by taking account
into a probe D4-brane moving in our D0-D4 background. The action for
a non-supersymmetric D4-brane is given as,

\begin{equation}
S_{D_{4}}=-\mu_{4}\int d^{5}xe^{-\phi}\mathrm{STr}\sqrt{-\det\left(g_{(5)}+\mathcal{F}\right)}+\frac{1}{2}\mu_{4}\mathrm{Tr}\int C_{1}\wedge\mathcal{F}\wedge\mathcal{F},\label{eq:23}
\end{equation}
where respectively $\mu_{4}=\frac{1}{\left(2\pi\right)^{4}l_{s}^{5}},g_{\left(5\right)},\mathcal{F}=2\pi\alpha^{\prime}F$
is the charge of the D4-brane, induced 5d metric and the gauge field
strength exited on the D4-brane. We assume that the non-vanished components
of $F$ are $F_{\mu\nu}\left(x\right)\delta^{1/2}\left(x^{4}-\bar{x}\right)$.
Then considering the $x^{4}$ direction is compacted on a circle $S^{1}$
with the period $\beta_{4}$, the action (\ref{eq:23}) can be expanded
in powers of $\mathcal{F}$ as a 4d Yang-Mills theory with a $\theta$
term,

\begin{equation}
S_{D_{4}}\simeq-\frac{1}{2g_{YM}^{2}}\mathrm{Tr}\int F\wedge^{*}F+i\frac{\theta}{8\pi^{2}}\mathrm{Tr}\int F\wedge F+\mathcal{O}\left(F^{3}\right),\label{eq:24}
\end{equation}
where the delta function is normalized as $\beta_{4}=\int dx^{4}\delta\left(x^{4}-\bar{x}\right)$
and the coupling constant $g_{YM},\theta$ are defined as,

\begin{align}
g_{YM}^{2}\left(U\right) & =\left[\mu_{4}\left(2\pi\alpha^{\prime}\right)^{2}\beta_{4}e^{-\phi}\sqrt{g_{44}}\right]^{-1}=\frac{8\pi^{2}g_{s}l_{s}}{\beta_{4}}\cosh\left[\frac{q_{\theta}}{2g_{s}}\rho\left(U\right)\right],\nonumber \\
\theta\left(U\right) & =-\frac{i}{l_{s}}\int_{\partial D=S_{x^{4}}^{1}}C_{1}=-\frac{i}{l_{s}}\int_{D}F_{2}=\theta-\frac{\beta_{4}}{g_{s}l_{s}}\tanh\left[\frac{q_{\theta}}{2g_{s}}\rho\left(U\right)\right],\label{eq:25}
\end{align}
which are the running couplings. Since the asymptotic region of the
bulk supergravity corresponds to the dual field theory, at the boundary
$\rho\rightarrow0,U\rightarrow\infty$ the Eq.(\ref{eq:25}) defines
the value of the Yang-Mills coupling constant and the $\theta$ angle
in the dual theory. In the large $N_{c}$ limit, we should define
the limitation $\bar{\theta}=\theta/N_{c}$ \cite{key-1,key-2} and
the t'Hooft coupling,

\begin{equation}
\lambda_{4}\left(U\right)=\frac{8\pi^{2}g_{s}l_{s}N_{c}}{\beta_{4}}\cosh\left[\frac{q_{\theta}}{2g_{s}}\rho\left(U\right)\right].\label{eq:26}
\end{equation}
According to the AdS/CFT dictionary, we remarkably find the Yang-Mills
and t'Hooft coupling constant $g_{YM},\lambda_{4}$ increase in the
IR region ($\rho\rightarrow\infty,U\rightarrow U_{T}$) while they
become small in the UV region ($\rho\rightarrow0,U\rightarrow\infty$)
with our D0-D4 solution. And this behavior is in qualitative agreement
with the property of asymptotic freedom in QCD or Yang-Mills theory.

To close this subsection, let us simply evaluate the relation of $q_{\theta}$
and $\theta$. In the Dp-brane supergravity solution, the normalization
of the Ramond-Ramond field $F_{p+2}$ is given as $2k_{0}^{2}\mu_{p}N_{p}=\int_{S^{8-p}}{}^{*}F_{p+2}$
and this normalization with (\ref{eq:7}) would tell us that $q_{\theta}$
is proportional to the number of D0-branes. Hence we have $q_{\theta}\sim g_{s}N_{0},N_{0}=g_{s}d_{\mathrm{D}_{0}}V_{4}$,
where $d_{\mathrm{D}_{0}}$ is the number density of D0-branes and
$V_{4}\simeq\left(2\pi R\right)^{3}\beta_{T}$ is the worldvolume
of the D4-branes. In order to include the influence of the D-instantons,
we further assume that $d_{\mathrm{D}_{0}}$ depends on $x^{4}$ because
$x^{4}=\theta R_{4}$ is periodic. This viewpoint implies that each
slice in the D4-brane with a fixed $x^{4}$ corresponds to a theta
vacuum in the dual field theory if we identify the coordinate $\theta$
to the theta angle in (\ref{eq:24}). So we could interpret that the
4d Yang-Mills action (\ref{eq:24}) is defined on a slice of the D4-brane
with $x^{4}=\bar{x}$, or namely with a theta angle $\theta=\bar{x}/R_{4}$
and it might offer a geometric interpretation of the theta-dependence
in the dual field theory. Finally we can define the dimensionless
density by using $\beta_{4}$ as $I\left(\theta\right)=d_{\mathrm{D}_{0}}\beta_{4}^{-4}$
which leads to $\left|q_{\theta}\right|\simeq2g_{s}V_{4}I\left(\theta\right)/\beta_{4}^{4}$.
Note that in the large $N_{c}$ limit $I\left(\theta\right)$ may
be expected to be a function of $\theta/N_{c}$.

\subsection{The thermodynamics}

The thermodynamics in holography is based on the relation between
the partition function of the bulk supergravity $Z_{\mathrm{SUGRA}}$
and the dual field theory (DFT) $Z_{\mathrm{DFT}}$ as $Z_{\mathrm{SUGRA}}=Z_{\mathrm{DFT}}$
in the large $N_{c}$ limit \cite{key-19,key-18,key-17}. Hence the
free energy density of the 4d theta-dependent Yang-Mills theory $f\left(\theta\right)$
could be obtained by

\begin{equation}
Z=e^{-V_{4}f\left(\theta\right)}=e^{-S_{\mathrm{SUGRA}}^{\mathrm{ren\ onshell}}},\label{eq:27}
\end{equation}
where $V_{4}$ and $S_{\mathrm{SUGRA}}^{\mathrm{ren\ onshell}}$ respectively
represents the 4d spacetime volume and the renormalized onshell action
of the bulk supergravity. For the duality to the thermal field theory,
$V_{4}=V_{3}\beta_{T}$ and $S_{\mathrm{SUGRA}}^{\mathrm{ren\ onshell}}$
respectively refers to its Euclidean version. The temperature in the
dual field theory is defined by $T=1/\beta_{T}$. So in order to avoid
the conical singularities in the dual field theory, it provides the
relation with our D0-D4 solution\footnote{It is not very obvious to find a relation as (\ref{eq:28}) just by
requiring no singularities outside the horizon with our gravitational
solution in Section 2.2. So we assume that our solution could return
to (\ref{eq:17}) continuously if $q_{\theta}\rightarrow0$ then we
find the relation (\ref{eq:28}) is at least valid at order $\mathcal{O}\left(q_{\theta}^{1}\right)$.},

\begin{equation}
2\pi T\simeq\left(\frac{3}{2}+\frac{q_{\theta}}{3ag_{s}}\right)\frac{U_{T}^{1/2}}{R^{3/2}}+\mathcal{O}\left(q_{\theta}^{2}\right),\label{eq:28}
\end{equation}
Afterwards the renormalized Euclidean onshell action of the supergravity
is given as,

\begin{equation}
S_{\mathrm{SUGRA}}^{\mathrm{ren\ onshell}}=S_{IIA}^{E}+S_{GH}+S_{CT},
\end{equation}
where $S_{IIA}^{E}$ refers to the Euclidean version of IIA supergravity
action (\ref{eq:3}) and $S_{GH},S_{CT}$ refers to the associated
Gibbons-Hawking and the bulk counter-term which are respectively given
as \cite{key-29,key-53},

\begin{align}
S_{IIA}^{E} & =-\frac{1}{2k_{0}^{2}}\int d^{10}x\sqrt{g}\left[e^{-2\phi}\left(\mathcal{R}+4\left(\partial\phi\right)^{2}\right)-\frac{1}{2}\left|F_{2}\right|^{2}-\frac{1}{2}\left|F_{4}\right|^{2}\right],\nonumber \\
S_{GH} & =-\frac{1}{k_{0}^{2}}\int d^{9}x\sqrt{h}e^{-2\phi}K,\nonumber \\
S_{CT} & =\frac{1}{k_{0}^{2}}\left(\frac{g_{s}^{1/3}}{R}\right)\int d^{9}x\sqrt{h}\frac{5}{2}e^{-7\phi/3}.\label{eq:30}
\end{align}
Here $h$ is the determinant of the boundary metric i.e. the slice
of the bulk metric (\ref{eq:4}) at fixed $\rho=\varepsilon$ with
$\varepsilon\rightarrow0$. And $K$ is the trace of the extrinsic
curvature at the boundary which is defined as,

\begin{equation}
K=\frac{1}{\sqrt{g}}\partial_{\rho}\left(\frac{\sqrt{g}}{\sqrt{g_{\rho\rho}}}\right)\bigg|_{\rho=\varepsilon}.
\end{equation}
Then the actions in (\ref{eq:30}) can be evaluated by using the D0-D4
solution discussed in Section 2.2. After some straightforward but
messy calculations, we finally obtain,

\begin{align}
S_{IIA}^{E} & =\mathcal{V}\left[\frac{3}{2\varepsilon}-\frac{9}{4}a+\frac{7q_{\theta}}{2g_{s}}\right],\nonumber \\
S_{GH} & =\mathcal{V}\left[-\frac{19}{6\varepsilon}+\frac{7}{6}\frac{9a^{2}g_{s}^{2}-36ag_{s}q_{\theta}+4q_{\theta}^{2}}{6ag_{s}^{2}+4g_{s}q}\right],\nonumber \\
S_{CT} & =\mathcal{V}\frac{5}{3\varepsilon},\label{eq:32}
\end{align}
and the free energy density $f\left(\theta\right)$ is therefore obtained
by using (\ref{eq:27}) (\ref{eq:32}) with the relation of $q_{\theta}$
and $\theta$ which is calculated as,

\begin{equation}
f\left(\theta,T\right)=-\frac{128N_{c}^{2}\pi^{4}T^{6}\lambda_{4}}{2187M_{KK}^{2}}+\frac{2M_{KK}^{5}\lambda_{4}}{3\pi^{2}T}I\left(\theta\right),\label{eq:33}
\end{equation}
where we have defined the Kaluza-Klein (KK) mass $M_{KK}=2\pi/\beta_{4}$
and rescaled $I\left(\theta\right)\rightarrow\left(2\pi l_{s}\right)^{3}M_{KK}^{3}I\left(\theta\right)$.
Besides the function $I\left(\theta\right)$ is turned out to be a
periodic and even function of $\theta$ i.e. $I\left(\theta\right)=I\left(-\theta\right),I\left(\theta\right)=I\left(\theta+2k\pi\right),k\in\mathbb{Z}$
and the energy of the true vacuum $F\left(\theta\right)$ is obtained
by minimizing the expression in (\ref{eq:33}) over $k$, 
\begin{equation}
F\left(\theta,T\right)=\mathrm{min}_{k}f\left(\theta,T\right).
\end{equation}
While at finite temperature the exact theta-dependence of the ground-state
free energy in Yang-Mills theory is less clear especially in the large
$N_{c}$ limit, the computation for one-loop contribution of instantons
to the functional integral at sufficiently high temperature suggests
that $f\left(\theta\right)-f\left(0\right)\propto1-\cos\theta$ \cite{key-1}.
Although this theta-dependence is consistent with the gravitational
constraints discussed in Section 2 i.e. $q_{\theta}\rightarrow0$
if $\theta\rightarrow0$, it does not have a definite limitation at
$N_{c}\rightarrow\infty$. Nonetheless if we assume the function $I\left(\theta\right)$
has a limit at $N_{c}\rightarrow\infty$, the topological susceptibility
can be computed by expanding (\ref{eq:33}) in powers of $\bar{\theta}$
as,

\begin{equation}
f\left(\bar{\theta}\right)-f\left(0\right)=\frac{2M_{KK}^{5}\lambda_{4}}{3\pi^{2}T}\sum_{n=1}^{\infty}\frac{b_{n}}{2n!}\bar{\theta}^{2n},\ b_{n}=\frac{\partial^{n}f\left(\bar{\theta},T\right)}{\partial\bar{\theta}^{n}}\bigg|_{\bar{\theta}=0}.
\end{equation}
Thus the topological susceptibility reads \footnote{Here the reader should notice the relation of $\bar{\theta}$ and
$\theta$ as $\bar{\theta}=\theta/N_{c}$ in the formulas.},
\begin{equation}
\chi\left(T\right)=\frac{\partial^{2}f\left(\bar{\theta},T\right)}{\partial\theta^{2}}\bigg|_{\theta=0}=\frac{2M_{KK}^{5}\lambda_{4}}{3\pi^{2}N_{c}^{2}T}b_{2}.\label{eq:36}
\end{equation}
where $b_{2}$ should be a positive numerical number\footnote{The function $I\left(\theta\right)$ may be expected as $I\left(\bar{\theta}\right)=1-\cos\bar{\theta}$
in the large $N_{c}$ limit and in this case the topological susceptibility
is $\chi\left(T\right)=\frac{2M_{KK}^{5}\lambda_{4}}{3\pi^{2}N_{c}^{2}T}b_{2}$
with $b_{2}=1/2$.}. The topological susceptibility (\ref{eq:36}) depends on temperature
as expected while it becomes vanished in the large $N_{c}$ limit.
We notice this large $N_{c}$ behavior remarkably agrees with the
simulation results reviewed in \cite{key-1} which indicates that
the topological susceptibility has a vanishing large $N_{c}$ above
the deconfinement temperature.

\section{Summary and discussion}

In this letter we holographically combine the IIA supergravity with
the theta-dependent Yang-Mills theory at finite temperature. The bulk
geometry is sourced by a stack of $N_{c}$ black D4-branes and $N_{0}$
D0-branes as D-instantons. As it is known in the pure black D4-brane
solution, the dual field theory indicates deconfinement at finite
temperature and adding D-instantons to the D4 background could describe
the dynamics of the theta angle in the bulk. To keep this duality
picture and include the dynamics of the D-instantons, we therefore
consider a sufficiently small backreaction from the D-instantons to
the bulk geometry, then a particular solution is found by solving
the IIA supergravity action. Afterwards using our supergravity solution,
we investigate the coupling constant and the ground-state energy as
two most fundamental properties in the dual field theory. The behavior
of the coupling constant exhibits the asymptotic freedom as in QCD
or Yang-Mills theory and the theta contribution to the free energy
density is suppressed at high temperature. The topological susceptibility
is vanished in the large $N_{c}$ limit. Remarkably all these results
are in qualitative agreement with various simulation results of the
theta-dependent Yang-Mills theory at finite temperature. And we in
addition propose a geometric interpretation of the theta-dependence
in this system.

In our D0-D4 background, the dual theory should deconfine at the temperature
$T\geq T_{c}$ where $T_{c}$ refers to the critical temperature of
the deconfinement transition. Below $T_{c}$ the current supergravity
solution would be invalid and the confinement in the dual theory should
be described by the bubble D0-D4 background as discussed in \cite{key-42,key-43,key-44,key-45,key-46,key-47,key-48,key-49,key-50}.
Notice that the thermodynamical variables have different large $N_{c}$
limits in these two D0-D4 backgrounds. The $T_{c}$ could be obtained
by comparing the free energy of our black (\ref{eq:33}) and the bubble
D0-D4 system \cite{key-42,key-43,key-44,key-45}, however the result
will remain substantially unchanged as it is given in \cite{key-45}
in the large $N_{c}$ limit. Another noteworthy point is that Eq.(\ref{eq:36})
implies the instantons would be more unstable in the dual theory at
high temperature due to the definition of the topological susceptibility
in QFT $\chi=-i\int d^{4}x\left\langle O\left(x\right)O\left(0\right)\right\rangle $,
where $O\left(x\right)=\mathrm{Tr}F\wedge F$ is the glueball condensate
operator. In other words, at extremely high temperature $T\gg T_{c}$
the quantum fluctuations would destroy the glueball condensate in
the dual theory in a very short time and the theta vacuum in the dual
field theory decays soon to the true vacuum. This conclusion is basically
consistent with e.g. \cite{key-7,key-8,key-9} and the D3-D(-1) approach
in \cite{key-52}.

To finish this paper, let us give the final comments. Despite our
holographic interpretation of the theta-dependence, the exact thermodynamics
involving the theta angle is still challenging both in gauge-gravity
duality and QFT, especially at finite temperature. In our theory,
this is reflected in that the specific relation of $q_{\theta}$ and
$\theta$ could not be determined naturally through the holographic
duality, thus we have to further require the density of the D0-branes
exactly controls the ground-state energy as the role of the theta
parameter in the dual field theory. While this could consistently
figure out the problem as we have done, the physical understanding
of this constraint is not clear. And unfortunately, the analysis in
QFT has not implied anything constructive yet, so we have to treat
it as a particular constraint in this system and leave it to the future
study.

\section*{Acknowledgements}

I would like to thank Wenhe Cai and Chao Wu for valuable comments
and discussions. S. W. L. is supported by the research startup funds
of Dalian Maritime University under Grant No. 02502608 and the Fundamental
Research Funds for the Central Universities under Grant No. 017192608.

\section*{Appendix: The D0-D4 solution in the $U$ coordinate}

We summarize the D0-D4 solution discussed in Section 2.2 here in the
$U$ coordinate. The components of the metric are written as\index{Commands!T!tag@\textbackslash tag},

\begin{equation}
ds^{2}=g_{\mu\nu}dx^{\mu}dx^{\nu}+g_{44}\left(dx^{4}\right)^{2}+g_{UU}dU^{2}+g_{\Omega\Omega}d\Omega_{4}^{2},\tag{A-1}\label{eq:37}
\end{equation}
where

\begin{align}
g_{00} & =-\left(\frac{U_{T}}{R}\right)^{3/2}\frac{f_{T}\left(U\right)^{\frac{9-4Q^{2}}{9+4Q^{2}}}}{\sqrt{2}}g_{1}\left(U\right)^{1/2}g_{2}\left(U\right)^{-1/2},\nonumber \\
g_{ij} & =\frac{1}{\sqrt{2}}\left(\frac{U_{T}}{R}\right)^{3/2}g_{1}\left(U\right)^{1/2}g_{2}\left(U\right)^{-1/2}\delta_{ij},\nonumber \\
g_{44} & =\left(\frac{U_{T}}{R}\right)^{3/2}\sqrt{2}f_{T}\left(U\right)^{\frac{12Q}{9+4Q^{2}}}\left[g_{1}\left(U\right)g_{2}\left(U\right)\right]^{-1/2},\nonumber \\
g_{UU} & =\left(\frac{9+4Q^{2}}{9+6Q}\right)^{2/3}\left(\frac{R}{U_{T}}\right)^{3/2}\frac{\left[g_{1}\left(U\right)g_{2}\left(U\right)\right]^{1/2}}{\sqrt{2}f_{T}\left(U\right)},\nonumber \\
g_{\Omega\Omega} & =\left(\frac{9+4Q^{2}}{9+6Q}\right)^{2/3}\left(\frac{R}{U_{T}}\right)^{3/2}\frac{U^{2}}{\sqrt{2}}\left[g_{1}\left(U\right)g_{2}\left(U\right)\right]^{1/2},\tag{A-2}\label{eq:38}
\end{align}
and the dilaton is

\begin{equation}
e^{\phi}=g_{s}\left(\frac{U_{T}}{R}\right)^{3/4}f_{T}\left(U\right)^{\frac{Q\left(3-2Q\right)}{9+4Q^{2}}}\frac{g_{1}\left(U\right)^{3/4}g_{2}\left(U\right)^{-1/4}}{2^{3/4}}.\tag{A-3}\label{eq:39}
\end{equation}
The parameter $Q$ and functions $g_{1,2}$ are defined as,

\begin{equation}
g_{1}\left(U\right)=1+f_{T}\left(U\right)^{\frac{2Q\left(3+2Q\right)}{9+4Q^{2}}},g_{2}\left(U\right)=1-f_{T}\left(U\right)^{\frac{9+6Q}{9+4Q^{2}}},Q=\frac{\left|q_{\theta}\right|}{ag_{s}}.\tag{A-4}
\end{equation}
Note that $Q$ is a positive number and if it is sufficiently small
we have $f_{T}\left(U\right)^{\frac{12Q}{9+4Q^{2}}}\simeq1,g_{1}\left(U\right)\simeq2$
in the region $U\in\left(U_{T}+\varepsilon,\infty\right)$ where $\varepsilon\rightarrow0$.
The metric (\ref{eq:38}) and the dilaton (\ref{eq:39}) returns to
the zero-th order solution (\ref{eq:17}) consistently if we set $q_{\theta},Q=0$.

\end{document}